\documentclass[11pt]{article}

\usepackage{amsmath,amssymb,bm}
\usepackage{physics}
\usepackage{graphicx}
\usepackage{lmodern}
\usepackage[T1]{fontenc}
\usepackage[utf8]{inputenc}
\usepackage{microtype}
\usepackage{hyperref}
\usepackage{xcolor}
\usepackage[margin=2.5cm]{geometry}
\usepackage{abstract}

\graphicspath{{./}}
\usepackage{authblk}

\usepackage{soul}

\usepackage{bm}
\newcommand{\bx}{\boldsymbol{x}}
\newcommand{\by}{\boldsymbol{y}}

\begin{document}

\title{Entanglement generation in a two-body Schr\"odinger--Newton model}

\author[1,2]{Marcin P{\l}odzie\'n}
\author[3,4]{Julia Os\k{e}ka-Lenart}
\author[2]{Maciej Lewenstein}
\author[5]{Micha{\l} Eckstein}

\affil[1]{Qilimanjaro Quantum Tech, Carrer de Vene\k{c}uela 74, 08019 Barcelona, Spain}
\affil[2]{ICFO-Institut de Ciencies Fotoniques, The Barcelona Institute of Science and Technology, 08860 Castelldefels (Barcelona), Spain}
\affil[3]{Astronomical Observatory, Faculty of Physics, Astronomy and Applied Computer Science Jagiellonian University, Orla 171, Krak\'ow, 30-244, Poland}
\affil[4]{Doctoral School of Exact and Natural Sciences, Jagiellonian University, ul.\ {\L}ojasiewicza 11, 30--348 Krak\'ow, Poland}
\affil[5]{Institute of Theoretical Physics, Faculty of Physics, Astronomy and Applied Computer Science, Jagiellonian University, ul.\ {\L}ojasiewicza 11, 30--348 Krak\'ow, Poland}

\maketitle

\begin{abstract}
The Schr\"odinger--Newton (SN) equation provides a semiclassical framework for the evolution of self-gravitating of massive quantum systems. We propose a two-body Schr\"odinger--Newton model that separates local nonlinear self-localization from the nonseparable Newtonian pair potential. Analytically, we show that the nonlinear self-field preserves the Schmidt spectrum, whereas direct entanglement generation arises from the nonseparable pair potential. Using numerical simulations in a regularized one-dimensional geometry, we find that entanglement generation depends sensitively on the initial spatial configuration and on the mass ratio. Highly localized, self-bound wavepackets experience minimal entanglement growth during scattering. Spatial delocalization and kinetic dispersion broaden the interaction region, amplifying the entangling power of the pair potential and exciting higher-order spatial modes. For dispersive Gaussian initial states, mass asymmetry shatters the lighter particle, producing Wigner negativity and rapid entanglement growth, whereas stationary SN profiles strongly suppress this effect. Stationary SN profiles isolate the bare pair-potential contribution; dispersive Gaussian initial states inflate it.
\end{abstract}

\section{Introduction}
\label{sec:introduction}

The Schr\"odinger--Newton (SN) equation aims at modelling the
self-gravitational evolution of a quantum wavefunction.
In the single-particle case, the Newtonian potential is sourced by the
particle's own probability density, and the resulting nonlinear feedback
produces gravitational self-localization, whereby a spreading wavepacket
is re-focused by its self-generated gravitational well.
This mechanism has been studied extensively as a candidate for
explaining the quantum-to-classical transition through gravity
\cite{Diosi1984,Penrose1996,Moroz1998,Tod1999,Giulini2011,Meter2011,Manfredi2013,Giulini2013,Anastopoulos2014,Bahrami2014,DunajskiPenrose23,Osekat2025}. The single-particle SN equation can be derived from different principles: coupled Schr\"odinger--Poisson equations \cite{Penrose96}, the non-relativistic limit of Einstein--Dirac or Einstein--Klein--Gordon systems \cite{GiuliniGrossardt2014},  semi-classical gravity \cite{Diosi87,Bahrami2014} or a Hartree approximation to fully quantum gravity \cite{Bahrami2014}. Independently, the SN equation serves as an effective model to describe Cold Dark Matter \cite{Schr_Poiss_Vlas_Poiss,2024PhRvR_VP_SP,2025arXiv_fluid_analog_SN}, boson stars evolution \cite{boson_stars_NatCom16} or interactions in ultra cold gases \cite{PhysRevLett.84.5687,PhysRevA.63.031603,Giovanazzi_2001,PhysRevA.65.053616,PhysRevA.76.053604,PhysRevA.78.013615}.

For more than one particle, the  prescription for the dynamics of self-gravitating quantum systems is no longer unique. One must specify
whether the classical gravitational field is sourced only through one-body marginal densities, through a shared mean
field, or through an explicit pair operator depending on $x_1-x_2$. These choices are
dynamically inequivalent. Marginal-density mean fields preserve product states, whereas a
nonseparable pair operator can change the Schmidt spectrum of an initially factorized
state. Entanglement-witness proposals aim to directly probe this distinction by testing
whether the interaction between two masses can increase their Schmidt rank
\cite{Bose2017,Marletto2017,Carney2019,Christodoulou2019,Belenchia2018,QG_RMP}.

The canonical $N$-particle Schr\"odinger--Newton equation is based on semiclassical mean-field gravity, $G_{\mu\nu} = 8\pi G/c^4 \langle \hat{T}_{\mu\nu} \rangle$, \`a la M\o ller--Rosenfeld \cite{Diosi1984,Bahrami2014}. In the Newtonian regime it yields separability-preserving mean-field interactions that preclude entanglement
\cite{Paterek2024}. In contrast, the standard nonrelativistic quantum treatment of two gravitating
masses contains a linear nonseparable Newtonian pair potential, analogous to the standard Coulomb electromagnetic interaction (see e.g. \cite{Coulomb}). Here we propose a two-body  Schr\"odinger--Newton-type equation that supplements nonlinear semiclassical self-localization with a nonseparable pairwise Newtonian
interaction. By including both mechanisms,
the present model isolates which term is responsible for entanglement and provides a direct
comparison against strictly separability-preserving mean-field reductions.

The proposed model is phenomenological rather than a first-principles relativistic theory.
However, any extension that combines Di\'osi--Penrose-type self-localization
\cite{Diosi1984,Penrose1996,Bahrami2014} with Newtonian mutual attraction must specify
whether the mutual interaction is a separability-preserving mean field or a nonseparable
pair potential. Formally separating these mechanisms allows us to identify which term
changes the Schmidt spectrum and which term only reshapes the one-body marginals.

We then solve the equation numerically for two-particle collisions in 1D.
The simulations demonstrate that entanglement growth is driven by the nonadditive Newtonian pair
potential. For dispersive Gaussian initial states, mass asymmetry shatters the lighter particle,
producing Wigner negativity and rapid entanglement growth. In contrast, in the localized-product mass-asymmetry scan, highly
localized stationary profiles strongly suppress this disruption and remain weakly entangled across
all simulated mass ratios.

The paper is organized as follows. Section~\ref{sec:model} introduces our two-body
Schr\"odinger--Newton model, its effective Hamiltonian structure, and the entanglement
measure. Section~\ref{sec:numerics} presents the numerical results,
introducing each initial state and its simulation in turn. Section~\ref{sec:conclusions} contains
the conclusions and an outlook into possible developments of realistic physical models.

\section{Two-Body Schr\"odinger--Newton Model}
\label{sec:model}

Our starting point is the Schr\"odinger--Newton equation \cite{Diosi1984,Penrose1996,Bahrami2014},
\begin{equation}\label{SN}
  i\hbar \, \partial_t \psi(t,\bx)  = \left[ -\frac{\hbar^2}{2m}\nabla^2-Gm^2\int\frac{|\psi(t,\by)|^2}{|\bx-\by|}d^3\by \right] \psi(t,\bx),
		\end{equation}
        modelling the time-evolution of the wavefunction of a single self-gravitating quantum particle of mass $m$.
We consider a 1D analogue of Eq. \eqref{SN} in dimensionless units (cf. \cite{Osekat2025})
\begin{equation}
i\,\partial_t\psi(x,t) = \left[ -\frac{1}{2\mu}\partial_x^2 - \kappa\mu^2 \int dy \, U_\epsilon(x-y) |\psi(y,t)|^2 \right] \psi(x,t),
\label{eq:single_SN}
\end{equation}
where $\kappa$ characterizes the self-gravitational coupling, and $U_\epsilon(r)$ is a one-dimensional softened inverse-distance kernel,
\begin{equation}
U_\epsilon(r)
=
\frac{1}{\sqrt{r^2+\epsilon^2}},
\label{eq:kernel}
\end{equation}
with $\epsilon>0$. This choice is a softened three-dimensional inverse-distance
interaction restricted to a one-dimensional line (cf. \cite{Coulomb}), avoiding the linear growth $|r|/2$ of
the strict one-dimensional Poisson Green's function. All attractive gravitational terms
are defined with explicit minus signs, with $U_\epsilon(r)>0$. The strictly
one-dimensional geometry enforces head-on encounters at zero impact parameter, which
enhances interaction effects relative to generic three-dimensional scattering. While the
quantitative peak values of the generated entanglement depend on the short-distance cutoff
$\epsilon$, the qualitative entanglement hierarchy we observe is primarily controlled by
phase-space overlap and initial spatial geometry.

To generalize this to a bipartite system, we treat the two masses as distinguishable subsystems with coordinates $x_1$ and $x_2$. We do not impose bosonic or fermionic exchange symmetry.
We propose the following two-body equation of motion:
\begin{equation}
i\,\partial_t\Psi(x_1,x_2,t)
=
\left[
-\frac{1}{2\mu_1}\partial_{x_1}^2
-\frac{1}{2\mu_2}\partial_{x_2}^2
-\kappa\mu_1\Phi_1(x_1,t)
-\kappa\mu_2\Phi_2(x_2,t)
-\gamma\mu_1\mu_2 U_\epsilon(x_1-x_2)
\right]\Psi(x_1,x_2,t),
\label{eq:pair_SN}
\end{equation}

with local gravitational self-potentials 
\begin{align}
\Phi_1(x_1,t) &= \int U_\epsilon(x_1-x_1')\,\mu_1\rho_1(x_1',t)\,dx_1', \label{eq:pair_phi1} \\[4pt]
\Phi_2(x_2,t) &= \int U_\epsilon(x_2-x_2')\,\mu_2\rho_2(x_2',t)\,dx_2', \label{eq:pair_phi2}
\end{align}
sourced dynamically by the marginal
densities,
\begin{equation}
\rho_1(x_1,t) = \int |\Psi(x_1,x_2,t)|^2\,dx_2,
\qquad
\rho_2(x_2,t) = \int |\Psi(x_1,x_2,t)|^2\,dx_1,
\label{eq:marginals}
\end{equation}
of the normalized two-particle wavefunction, $\int
dx_1\,dx_2\,|\Psi(x_1,x_2,t)|^2=1$. The direct pair potential is a standard linear two-body operator:
\begin{equation}
V_{\rm pair}(x_1,x_2) = -\gamma\mu_1\mu_2 U_\epsilon(x_1-x_2).
\label{eq:pair_potential}
\end{equation}

Here, the parameters $\kappa$ and $\gamma$ control the self-gravitational and
pair-gravitational couplings, respectively. While physically there is only one Newton
constant ($\kappa=\gamma$), we retain distinct symbols to distinguish nonlinear
self-energy renormalization ($\kappa$) from linear inter-particle pair interaction
($\gamma$). We set $\hbar=1$ and
measure lengths, masses, and times in units $x_0$, $m_0$, and $t_0=m_0 x_0^2/\hbar$, so
that the dimensionless masses are $\mu_i=m_i/m_0$ and the dimensionless couplings
$\kappa$, $\gamma$ absorb the physical Newton constant and the chosen regularization
scale.

The phenomenological equation of motion \eqref{eq:pair_SN} admits the following
conserved total energy functional:
\begin{equation}\label{eq:energy}
\begin{split}
E[\Psi] &= \int dx_1\,dx_2\, \Psi^* \left( T_1 + T_2 + V_{\rm pair} \right) \Psi \\
&\quad - \frac{\kappa}{2}\mu_1^2 \iint \rho_1(x) U_\epsilon(x-x') \rho_1(x')\,dx\,dx' \\
&\quad - \frac{\kappa}{2}\mu_2^2 \iint \rho_2(x) U_\epsilon(x-x') \rho_2(x')\,dx\,dx',
\end{split}
\end{equation}
where $T_i = -(2\mu_i)^{-1} \partial_{x_i}^2$ are the kinetic energy operators, and the factors of $1/2$ correctly prevent double-counting of the self-energy. Formula \eqref{eq:energy} shows that the effective generator of dynamics \eqref{eq:pair_SN} is a time- and state-dependent `Hamiltonian':
\begin{equation}
H_{\rm eff}(t) = H_1[\rho_1](t) \otimes I_2 + I_1 \otimes H_2[\rho_2](t) + V_{\rm pair},
\end{equation}
where the nonlinear single-particle Schr\"odinger--Newton operators are
\begin{equation}
H_i[\rho_i](t) = -\frac{1}{2\mu_i}\partial_{x_i}^2 - \kappa\mu_i\Phi_i(x_i,t).
\end{equation}

The self-field terms $\Phi_i$ are nonlinear functionals of the state, but their
operator structure is local. Setting $\gamma=0$, the generator has the form
\[
H_{\rm self}(t)=A[\rho_1](t)\otimes I+I\otimes B[\rho_2](t),
\]
with $A$ and $B$ Hermitian along the solution. The reduced density operator (whose diagonal elements yield the spatial density $\rho_1(x_1) = \rho_1^{\rm red}(x_1, x_1)$) then obeys
\[
\partial_t\rho_1^{\rm red}
=
-i[A[\rho_1](t),\rho_1^{\rm red}],
\]
because the partial trace of the $B$ commutator vanishes. Hence, for any integer $n \ge 1$,
\[
\frac{d}{dt}\Tr[(\rho_1^{\rm red})^n]
=
-in\,\Tr\!\left[(\rho_1^{\rm red})^{n-1}
[A[\rho_1](t),\rho_1^{\rm red}]\right]
=0 .
\]
Thus the self-field flow strictly preserves the eigenvalue spectrum of the reduced density operator. It may
change the Schmidt modes, but it cannot change the Schmidt eigenvalues.

The direct pair term $V_{\rm pair}$ is linear in $\Psi$ but nonseparable as an operator on
$\mathcal{H}_1\otimes\mathcal{H}_2$. For the product initial states considered below, the residual nonadditive part
of this operator has nonzero covariance, and the pair term therefore changes
the Schmidt spectrum at order $t^2$.

To quantify this entanglement generation, we first write the two-body state in its Schmidt decomposition:
\begin{equation}
\Psi(x_1, x_2, t) = \sum_k \sqrt{\lambda_k(t)}\, u_k(x_1) v_k(x_2),
\label{eq:schmidt_decomp}
\end{equation}
where $u_k(x_1)$ and $v_k(x_2)$ are the orthonormal spatial Schmidt modes for each particle. The reduced density operator of particle 1 then has the kernel
\begin{equation}
\rho_1^{\rm red}(x_1,x_1';t)
=
\int \Psi(x_1,x_2,t)\Psi^*(x_1',x_2,t)\,dx_2
= \sum_k \lambda_k(t) u_k(x_1)u_k^*(x_1'),
\label{eq:reduced_density}
\end{equation}
with real, non-negative Schmidt eigenvalues $\lambda_k(t)$ satisfying $\sum_k \lambda_k = 1$.

The von Neumann entanglement entropy reads
\begin{equation}
S_{\rm vN}(t)
=
-\Tr\left[
\rho_1^{\rm red}(t)\ln\rho_1^{\rm red}(t)
\right]
=
-\sum_k \lambda_k(t)\ln \lambda_k(t).
\label{eq:vN_entropy}
\end{equation}

For a product initial state at $t = 0$, $\Psi(x_1, x_2, 0) = \psi_1(x_1, 0) \psi_2(x_2, 0)$, the
reduced density operator is a rank-one projector $\rho_1^{\rm red}(x_1, x_1'; 0) = \psi_1(x_1, 0) \psi_1^*(x_1', 0)$,
and therefore the initial entanglement is zero, $S_{\rm vN}(0) = 0$.

For product initial states, separability-preserving mean-field projections identically yield $S_{\rm vN}(t)=0$. This effect was also observed for the canonical 2-particle Schr\'odinger--Newton equation \cite{Paterek2024} and is analyzed in full generality in Appendix \ref{sec:appendix_hartree}.In contrast, the
nonseparable pair interaction generically yields $S_{\rm vN}(t) > 0$.

For smooth product initial states $\Psi(x_1,x_2,0) = \psi_1(x_1,0)\psi_2(x_2,0)$, the
linear entropy $S_L(t) = 1 - \mathrm{Tr}[\rho_1^{\rm red}(t)^2]$ vanishes at $t=0$
together with its first derivative. The leading-order entanglement generation is therefore
quadratic in time:
\begin{equation}
S_L(t) = 2t^2\left[\mathbb{V}_{12}\!\left(V_{\rm pair}\right) - \mathbb{V}_{1}\!\left(\langle V_{\rm pair}\rangle_2\right) - \mathbb{V}_{2}\!\left(\langle V_{\rm pair}\rangle_1\right)\right] + O(t^3),
\label{eq:SL_secondorder}
\end{equation}
where $\mathbb{V}_{12}$ denotes the variance in the joint product state, and
$\mathbb{V}_{1}$, $\mathbb{V}_{2}$ denote variances over the respective marginals.
Explicitly, these are defined as:
\begin{align}
\mathbb{V}_{12}(V_{\rm pair}) &= \langle V_{\rm pair}^2 \rangle_{12} - \langle V_{\rm pair} \rangle_{12}^2, \\
\mathbb{V}_{1}(\langle V_{\rm pair}\rangle_2) &= \langle \langle V_{\rm pair}\rangle_2^2 \rangle_{1} - \langle \langle V_{\rm pair}\rangle_2 \rangle_{1}^2, \\
\langle V_{\rm pair}\rangle_2(x_1) &= \int |\psi_2(x_2)|^2 V_{\rm pair}(x_1,x_2)\,dx_2.
\end{align}
Here, $\langle V_{\rm pair}\rangle_{1,2}$ denote partial expectations over the respective
subsystems, with the symmetric definition
\begin{equation}
\langle V_{\rm pair}\rangle_1(x_2)
=
\int |\psi_1(x_1)|^2 V_{\rm pair}(x_1,x_2)\,dx_1.
\end{equation}
Defining the nonadditive residual interaction
\begin{equation}
V_{\rm res}(x_1,x_2)
=
V_{\rm pair}(x_1,x_2)
-
\langle V_{\rm pair}\rangle_2(x_1)
-
\langle V_{\rm pair}\rangle_1(x_2)
+
\langle V_{\rm pair}\rangle_{12},
\end{equation}
one may write
\begin{equation}
S_L(t)
=
2t^2\langle V_{\rm res}^2\rangle_{12}
+
O(t^3).
\label{eq:SL_residual}
\end{equation}
The coefficient is manifestly non-negative and vanishes when $V_{\rm pair}$ is additive on
the support of the initial product state, $V_{\rm pair}(x_1,x_2)=a(x_1)+b(x_2)$, since
then $V_{\rm res}=0$ and purely local contributions to the interaction do not generate
entanglement. We use the linear entropy $S_L$ for this short-time expansion to avoid $t^2
\ln t$ non-analyticities associated with the von Neumann entropy.

\subsection{Phase-space Wigner diagnostics}
\label{sec:wigner_def}

The entanglement diagnostics defined above are based on the Schmidt decomposition, which
captures mode occupations but provides no spatial or momentum resolution. We use reduced
Wigner functions to distinguish simple squeezing from wavepacket interference and spatial
fragmentation.

For the reduced density operator $\hat{\rho}_1$ of particle~1 with position-space kernel
$\rho_1^{\rm red}(x,x')$ defined in Eq.~\eqref{eq:reduced_density}, the Wigner function is
\begin{equation}
W(x,p) = \frac{1}{\pi}\int dy\; \rho_1^{\rm red}\!\left(x+y,\, x-y\right)\, e^{-2ipy},
\label{eq:wigner_def}
\end{equation}
where $x$ and $p$ are the phase-space coordinates of particle~1, and the integral runs
over the displacement variable~$y$. This definition is normalized so that $\int dp\,
W(x,p) = \rho_1(x)$ and $\int dx\,dp\, W(x,p) = 1$. While pure Gaussian states possess strictly non-negative Wigner functions, mixed reduced states can exhibit Wigner negativity. Here we use $W(x,p)$ as a phase-space diagnostic of the local non-Gaussian structure associated with the Schmidt-spectrum evolution.

The single-particle reduced Wigner functions visualize the local phase-space state of each
subsystem, but they are insensitive to correlations between the two particles. A more
interaction-adapted diagnostic is obtained by tracing over the center-of-mass degree of freedom and Wigner-transforming the relative coordinate, since the pair potential depends only on $r=x_1-x_2$.

We introduce center-of-mass and relative coordinates for equal masses ($\mu_1=\mu_2$):
\begin{equation}
X_{\rm cm} = \tfrac{1}{2}(x_1+x_2), \qquad r = x_1 - x_2,
\label{eq:cm_rel_coords}
\end{equation}
with conjugate momenta $P = p_1+p_2$ (CM) and $p_{\rm rel} = \tfrac{1}{2}(p_1-p_2)$ (relative).
The full two-body Wigner function $W^{(2)}(X_{\rm cm},P,r,p_{\rm rel})$ is a
four-dimensional quasiprobability distribution. Marginalizing over the CM degree of
freedom defines the \emph{relative-coordinate reduced density operator}
\begin{equation}
\rho_{\rm rel}(r,r') = \int dX_{\rm cm}\;
\Psi\!\left(X_{\rm cm}+\tfrac{r}{2},\,X_{\rm cm}-\tfrac{r}{2}\right)
\Psi^*\!\left(X_{\rm cm}+\tfrac{r'}{2},\,X_{\rm cm}-\tfrac{r'}{2}\right),
\label{eq:rho_rel}
\end{equation}
and its Wigner transform
\begin{equation}
W_{\rm rel}(r,p_{\rm rel})
=
\frac{1}{\pi}\int ds\;
\rho_{\rm rel}(r+s,\,r-s)\,e^{-2ip_{\rm rel}s}
\label{eq:wigner_rel}
\end{equation}
is the marginal of $W^{(2)}$ over the CM phase space: $W_{\rm rel}(r,p_{\rm rel}) = \int
dX_{\rm cm}\,dP\;W^{(2)}(X_{\rm cm},P,r,p_{\rm rel})$. The normalization $\int dr\,dp_{\rm
rel}\,W_{\rm rel} = 1$ holds, and the marginal $\int dp_{\rm rel}\,W_{\rm rel}(r,p_{\rm
rel}) = \rho_{\rm rel}(r,r)$ gives the probability density of finding the two particles at
relative separation~$r$. Because $V_{\rm pair}$ acts only on $r$, $W_{\rm rel}(r,p_{\rm
rel})$ is the most direct phase-space diagnostic of the degree of freedom driven by the
entangling interaction. The generation of negativity in $W_{\rm rel}$ indicates
non-Gaussian phase-space fragmentation, which accompanies, but does not by itself
quantify, the growth of the Schmidt spectrum.

For unequal masses (Sec.~\ref{sec:mass_asymmetry}), we use the mass-weighted
center-of-mass coordinate $X_{\rm cm}=(\mu_1 x_1+\mu_2 x_2)/(\mu_1+\mu_2)$ with conjugate
momenta $P=p_1+p_2$ and $p_{\rm rel}=(\mu_2 p_1-\mu_1 p_2)/(\mu_1+\mu_2)$. The relative
coordinate remains $r=x_1-x_2$, and the inverse transformation is
\begin{equation}
x_1=X_{\rm cm}+\frac{\mu_2}{\mu_1+\mu_2}r,
\qquad
x_2=X_{\rm cm}-\frac{\mu_1}{\mu_1+\mu_2}r.
\end{equation}
For unequal masses, Eq.~\eqref{eq:rho_rel} is modified by using this inverse map inside
the wavefunction arguments.

This separation will be used throughout the simulations below: $\kappa$ changes the
marginal density profiles entering the scattering problem, whereas $\gamma$ supplies the
nonseparable operator that changes the Schmidt spectrum. In the Appendix, we show that
projecting the same pair-interaction model onto the Hartree product manifold recovers the
separability-preserving coupled mean-field structure of Ref.~\cite{Paterek2024}. Thus the
absence of entanglement generation in Hartree mean-field equations follows from the
product-state restriction rather than from the local SN self-field itself.

\section{Numerical results}
\label{sec:numerics}

The evolution is computed using second-order Strang splitting on a one-dimensional
periodic grid of length $L=40$; the nonlinear self-potentials are recomputed from the
instantaneous marginal densities at each potential substep. The gravitational potentials
are evaluated via fast Fourier transform convolutions. The convolution entering the
self-potentials is evaluated with the same periodic distance convention as the split-step
grid; the pair kernel $U_\epsilon(x_i-x_j)$ is evaluated using the same periodic
minimum-distance convention. The simulation domain is chosen large enough that the density
remains negligible near the boundaries over the reported time window. Direct checks with
$L=60$ confirmed that periodic-image effects do not affect the quoted entropy values at
the stated accuracy. We use $\epsilon=0.2$, which provides a physically sharp pair-scattering cutoff while remaining robustly resolved on the spatial grid ($dx \approx 0.156$).
The Schmidt spectrum is computed by performing a singular value decomposition (SVD) directly on the discretized two-particle wavefunction.

Unless stated otherwise, the baseline simulations use equal masses $\mu_1=\mu_2=1$,
initial mean separation $R_0 = |\langle x_1 \rangle - \langle x_2 \rangle|_{t=0} = 6$, width parameter $\sigma_0=1$, and a universal gravitational
coupling $\lambda \equiv \kappa = \gamma = 1$. Quoted peak entropies refer to the plotted
window $t\in[0,40]$ unless a longer time horizon is explicitly indicated.
The total energy $E_{\rm total}=E_K+E_{\rm int}+E_{\rm self}$ is monitored
throughout all runs. The relative energy drift satisfies $\max_t |E(t)-E(0)|/|E(0)| <
10^{-8}$, and the norm drift remained below $10^{-13}$. Convergence was verified by
doubling the grid ($N=256\to512$), extending the domain ($L=40\to60$), and halving the
time step ($\Delta t=0.01\to0.005$); the peak entropy values changed by less than
$10^{-4}$ in absolute terms, and final separations shifted by less than $10^{-3}$. The
quoted entropy values are therefore insensitive to the grid, domain size, and time step at
the precision reported here.

We use two interparticle separation diagnostics. For the localized-product collision, we
report the marginal center-of-mass separation
\begin{equation}
\Delta x_{\rm mean}(t) \equiv |\langle x_1\rangle_t - \langle x_2\rangle_t|.
\label{eq:dx_cm}
\end{equation}
For superposition states whose marginals can be symmetric by construction, we additionally
report the rms relative separation
\begin{equation}
D_{\rm rel}(t) = \sqrt{\langle (x_1-x_2)^2\rangle_t},
\label{eq:d_rel}
\end{equation}
which accurately measures the true interparticle distance even when the state is spatially
delocalized over multiple distinct regions, a feature that the simple marginal difference
$\Delta x_{\rm mean}$ cannot resolve. Here, $D_{\rm rel}$ represents the true quantum mechanical observable for the root-mean-square distance.

\subsection{Initial state configurations}
\label{sec:initial_states}

We prepare each initial wavepacket either as a Gaussian or as the stationary ground state
of the single-particle SN equation.

The stationary single-particle SN profiles used as initial data are obtained by solving
the nonlinear eigenproblem
\begin{equation}
\omega\,\phi_{\rm SN}(x) = \left[-\frac{1}{2\mu}\partial_x^2 - \kappa\mu^2 \int U_\epsilon(x-x')\,|\phi_{\rm SN}(x')|^2\,dx'\right]\phi_{\rm SN}(x),
\label{eq:sn_eigenproblem}
\end{equation}
subject to $\int|\phi_{\rm SN}(x)|^2\,dx=1$. These states are obtained via imaginary-time
propagation starting from a Gaussian seed of width $\sigma_0$; for the ground-state component
used here, the converged profile is independent of the seed within numerical tolerance.

We construct the two-body initial state from these single-particle states in four distinct
configurations. The localized single-particle components \(\phi_L(x)\) and \(\phi_R(x)\)
are prepared by centering the chosen profile at \(-R_0/2\) and \(+R_0/2\), respectively.
For Gaussian wavepackets we use
\begin{equation}
\phi_{L/R}(x)\propto \exp[-(x\pm R_0/2)^2/(4\sigma_0^2)].
\end{equation}
For stationary-profile simulations, we obtain the single-particle ground state \(\phi_{\rm
SN}(x)\) by imaginary-time propagation and define the separated states by rigid
translation:
\begin{equation}
\phi_{L/R}(x)=\phi_{\rm SN}(x\pm R_0/2).
\end{equation}
Their initial spatial overlap is $s \equiv \int dx\,\phi_L^*(x)\phi_R(x)$, which for
Gaussians is $s = \exp[-R_0^2/8\sigma_0^2]$. For the real translated profiles used here, $s>0$.

Using these well-separated left ($L$) and right ($R$) spatial modes, we investigate the
following four initial two-body configurations:
\begin{align}
\Psi_{\rm I}(x_1,x_2) &= \phi_L(x_1)\phi_R(x_2), \label{eq:psi_I} \\
\Psi_{\rm II}(x_1,x_2) &= \mathcal{N}_{\rm II} [\phi_L(x_1)+\phi_R(x_1)][\phi_L(x_2)+\phi_R(x_2)], \label{eq:psi_II} \\
\Psi_{\rm III}(x_1,x_2) &= \mathcal{N}_{\rm III} [\phi_L(x_1)\phi_L(x_2)+\phi_R(x_1)\phi_R(x_2)], \label{eq:psi_III} \\
\Psi_{\rm IV}(x_1,x_2) &= \mathcal{N}_{\rm IV} [\phi_L(x_1)\phi_R(x_2)+\phi_R(x_1)\phi_L(x_2)], \label{eq:psi_IV}
\end{align}
where the normalizations are $\mathcal{N}_{\rm II} = [4(1+s)^2]^{-1/2}$ and
$\mathcal{N}_{\rm III} = \mathcal{N}_{\rm IV} = [2+2s^2]^{-1/2}$.

The localized product $\Psi_{\rm I}$ is purely separable ($S_{\rm vN}=0$). The delocalized
product $\Psi_{\rm II}$ is also separable, but for negligible overlap $s$, the marginal densities are split equally
between the two wells. The superpositions $\Psi_{\rm III}$ and $\Psi_{\rm IV}$ are
Bell-like entangled states; their two nonzero Schmidt eigenvalues are $\lambda_\pm = (1\pm
s)^2/[2(1+s^2)]$, giving $S_{\rm vN}(0) = -\lambda_+\ln\lambda_+ - \lambda_-\ln\lambda_-
\to \ln 2$ in the negligible-overlap limit $s\to 0$. $\Psi_{\rm III}$ features
position-correlated particles, while in $\Psi_{\rm IV}$ the particles are anti-correlated.

\subsection{Equal-mass collision dynamics}

In the following, we analyze the collision dynamics for equal masses ($\mu_1 = \mu_2 = 1$) across the
four initial configurations.

\subsubsection{Localized product}
\label{sec:product_collision}

The localized product state $\Psi_{\rm I}$ (Eq.~\ref{eq:psi_I}) has $S_{\rm vN}(0)=0$ and
a factorized joint probability density $|\Psi_{\rm I}|^2 = \rho_1(x_1)\rho_2(x_2)$
representing two wavepackets centered at opposite sides.

\begin{figure*}[htbp]
\centering
\includegraphics[width=\linewidth]{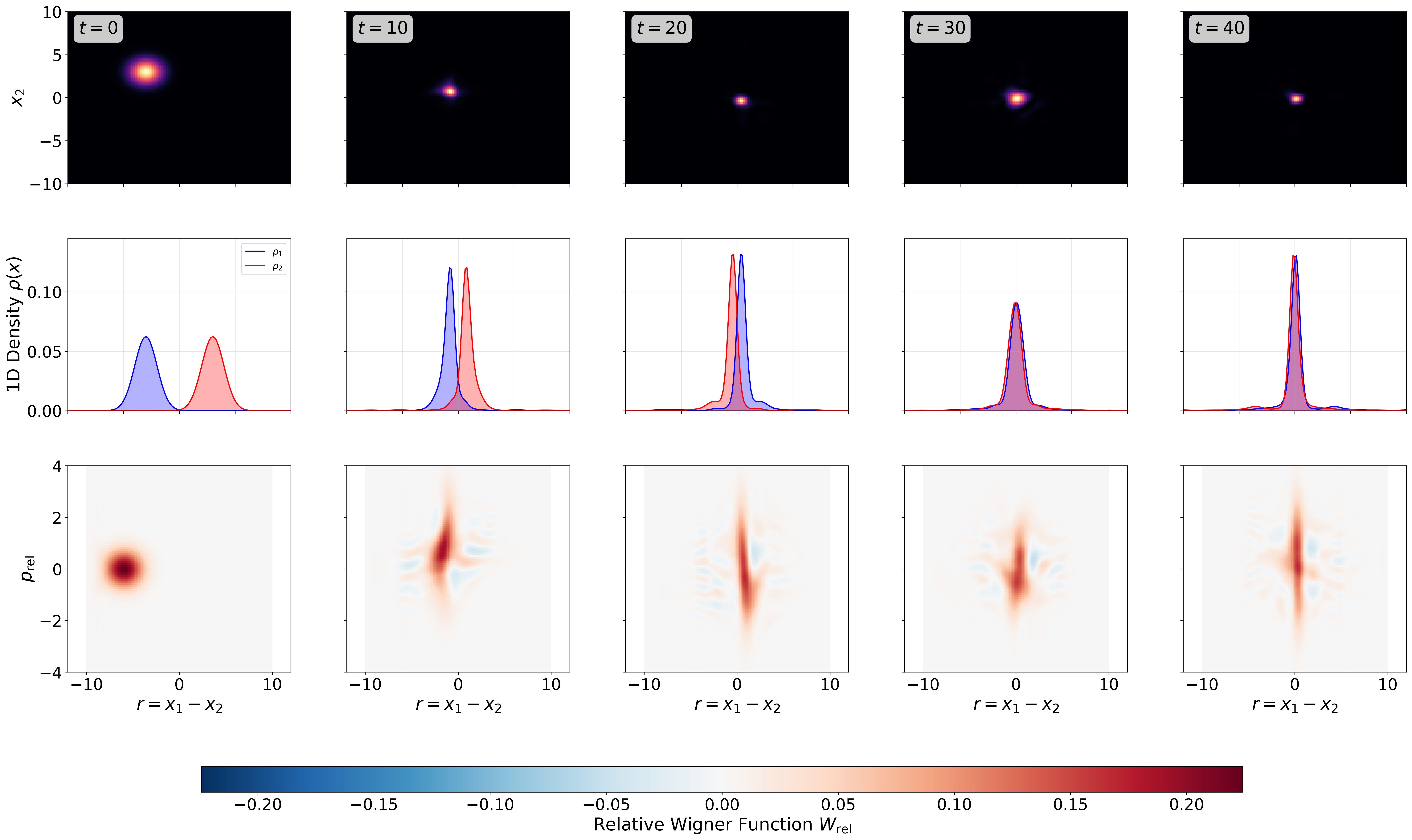}
\caption{
Collision dynamics for the localized product state $\Psi_{\rm I}$ (Eq.~\ref{eq:psi_I})
using Gaussian wavepackets. The snapshots are taken at $t=0, 10, 20, 30, 40$. The panel
shows (top row) the joint probability density $|\Psi(x_1,x_2)|^2$; (middle row)
single-particle marginal densities $\rho_1(x)$ (blue) and $\rho_2(x)$ (red); (bottom row)
the relative-coordinate Wigner function $W_{\rm rel}$.
Parameters: $R_0=6$, $\kappa=\gamma=1$, $\sigma_0=1$.
}
\label{fig:collision_product_gaussian}
\end{figure*}

\begin{figure*}[htbp]
\centering
\includegraphics[width=\linewidth]{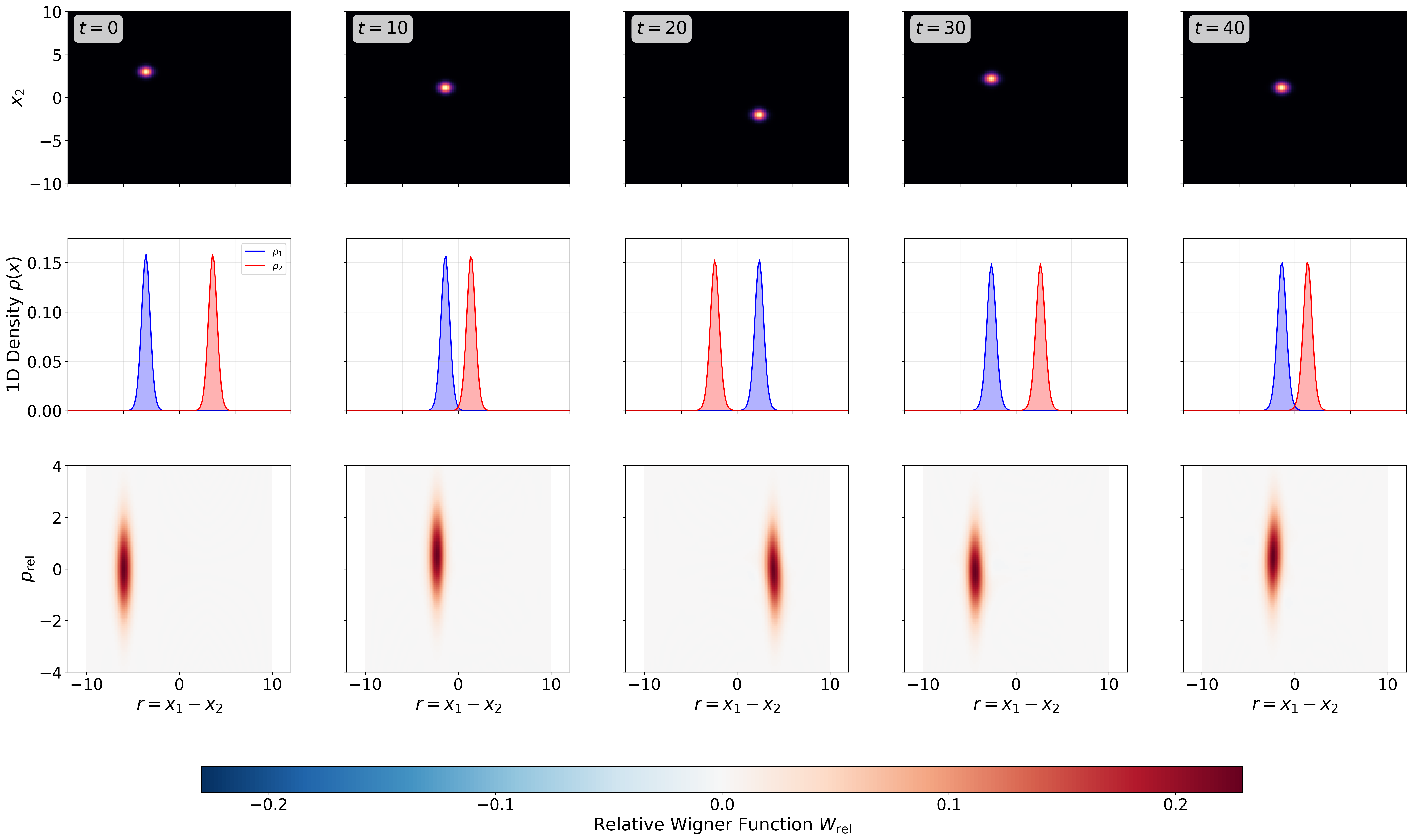}
\caption{
Collision dynamics for the localized product state $\Psi_{\rm I}$ (Eq.~\ref{eq:psi_I})
using stationary SN profiles. The snapshots are taken at $t=0, 10, 20, 30, 40$. The panel
shows (top row) the joint probability density $|\Psi(x_1,x_2)|^2$; (middle row)
single-particle marginal densities $\rho_1(x)$ (blue) and $\rho_2(x)$ (red); (bottom row)
the relative-coordinate Wigner function $W_{\rm rel}$.
Parameters: $R_0=6$, $\kappa=\gamma=1$; the imaginary-time seed width is $\sigma_0=1$.
}
\label{fig:collision_product_soliton}
\end{figure*}

The two wavepackets accelerate toward each other under mutual
gravitational attraction and collide.
The collision dynamics depend on the single-particle
profile (Figs.~\ref{fig:collision_product_gaussian} and~\ref{fig:collision_product_soliton}).

Gaussian wavepackets are not stationary states of the self-gravitational potential and
undergo kinetic dispersion. The broadened distributions overlap extensively, generating
entanglement ($S_{\rm vN} \approx 0.87$). After the collision, the Gaussians merge at the
center of mass ($\Delta x_{\rm mean} \to 0$), coalescing into a single central peak with
the entropy saturating at its post-collision value. The merger is effectively inelastic at
the level of the reduced one-body density: relative translational kinetic energy is
transferred into internal dispersive degrees of freedom, while total energy and norm
remain conserved.

Stationary SN profiles behave differently. Their self-bound structure is perturbed only
weakly during the collision. The particles remain gravitationally bound, oscillating
periodically in their mutual well. The entanglement per bounce is small ($S_{\rm vN}
\approx 0.19$ after a few collisions) but accumulates through repeated elastic scattering.
Extended simulations to $t=200$ confirm that the oscillation persists, with entanglement
saturating at $S_{\rm vN} \approx 0.29$.

At $t=0$, the relative-coordinate Wigner function $W_{\rm rel}$ (bottom rows of
Figs.~\ref{fig:collision_product_gaussian} and~\ref{fig:collision_product_soliton}) is a
single localized lobe centered at $r=-R_0$. For stationary profiles, it remains
predominantly positive throughout the collision, developing only mild phase-space shear.
For dispersive Gaussians, the broadening prior to collision produces pronounced
phase-space shearing and Wigner fragmentation.

\subsubsection{Delocalized product}
\label{sec:independent_collision}

\begin{figure}[htbp]
\centering
\includegraphics[width=\linewidth]{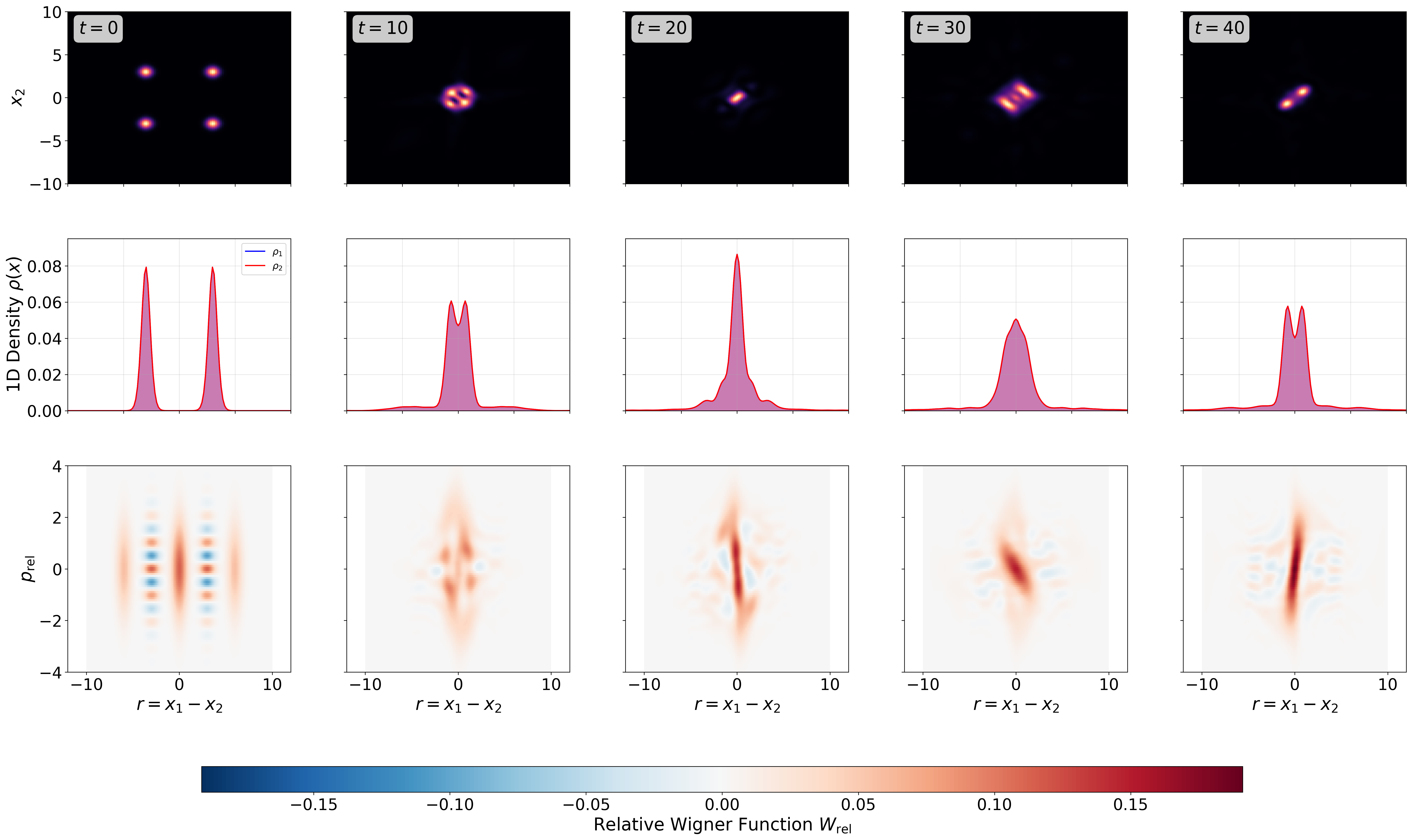}
\caption{
Evolution of the delocalized product state $\Psi_{\rm II}$ (Eq.~\ref{eq:psi_II}) using
stationary SN profiles.
Despite being initialized from stationary profiles, the LL/RR and LR/RL components acquire
different phases and accelerations under the pair potential, producing $S_{\rm
vN}\approx1.5$. The panel shows (top row) the joint probability density
$|\Psi(x_1,x_2)|^2$; (middle row) single-particle marginal densities $\rho_1(x)$ and
$\rho_2(x)$; (bottom row) the relative-coordinate Wigner function $W_{\rm rel}$, with
snapshots at $t=0, 10, 20, 30, 40$.
Parameters: $R_0=6$, $\kappa=\gamma=1$; the imaginary-time seed width is $\sigma_0=1$.
}
\label{fig:collision_independent}
\end{figure}

In the delocalized product configuration $\Psi_{\rm II}$ (Eq.~\ref{eq:psi_II}), each
particle is prepared in the same coherent superposition of left and right wave packets.
Although $\Psi_{\rm II}$ has $S_{\rm vN}(0)=0$, its joint density has four equal-weight
lobes corresponding to LL, LR, RL, and RR components (Fig.~\ref{fig:collision_independent}).

For negligible overlap $s$, the marginal density is split approximately equally between $L$ and $R$, so the local part of the self-gravitational well at each component is sourced by only half the total density, weakening the local
binding. The remote component generates a secondary attraction toward the center. The
delocalized product reaches $S_{\rm vN}\approx 1.58$, roughly an order of magnitude above
the localized-product collision ($S_{\rm vN}\approx 0.19$).

Although the local state for each particle initially spans a two-dimensional
subspace $\{\phi_L, \phi_R\}$, the observed entanglement $S_{\rm vN}\approx 1.58$ exceeds the rigid-profile maximum of $\ln 2 \approx 0.69$. This confirms that the full pair-interaction dynamics excites higher-order spatial modes well beyond the initial two-mode subspace.

The pair potential acts differently on the co-localized components ($\phi_L\otimes\phi_L$
and $\phi_R\otimes\phi_R$) and the separated components ($\phi_L\otimes\phi_R$ and
$\phi_R\otimes\phi_L$), because the former sample the smallest separations, where
$U_\epsilon(x_1-x_2)$ is largest and the attractive pair potential is most negative.

In phase space, the reduced Wigner function $W(x_1,p_1)$ initially describes a pure
coherent superposition $\psi_1(x) \propto \phi_L(x) + \phi_R(x)$,
containing two localized positive lobes centered at $(\pm R_0/2, 0)$ and interference
fringes between them. As the system evolves, the superposition-induced nonstationarity
sets in: because the marginal density is split, each lobe experiences only half the
self-gravitational binding depth. Simultaneously, the density of the remote spatial
component generates a secondary attractive potential toward the center. The lobes
accelerate inward, acquiring nonzero mean momenta $\langle p \rangle \neq 0$, visible as a
tilt of the Wigner lobes away from the $p=0$ axis.
During the collision, the two counter-propagating phase-space distributions overlap and
interfere. Small-scale interference fringes appear between the counter-propagating
components, and the generation of extended negative regions $\mathcal{V}_{-} = \int
dx\,dp\, |\min(W,0)|$ accompanies the growth of the entanglement entropy.

In the relative coordinate (bottom row of Fig.~\ref{fig:collision_independent}), the
initial state is a macroscopic superposition of co-localized ($r=0$) and separated ($r=\pm
R_0$) components, producing high-frequency interference fringes. The gravitational pair
potential pulls the separated components toward the center, shearing the relative phase
space and merging the components into a tilted central stripe. The resulting Wigner
negativity diagnoses non-Gaussian phase-space fragmentation.

\subsubsection{Correlated superposition}
\label{sec:corr_collision}

The correlated (``cat'') state $\Psi_{\rm III}$ (Eq.~\ref{eq:psi_III}) is a coherent
same-side superposition. Unlike the product states, this configuration is entangled at
$t=0$, with both particles localized on the same side ($L,L$ or $R,R$), but the
superposition cannot be factorized into single-particle states.

\begin{figure}[htbp]
\centering
\includegraphics[width=\linewidth]{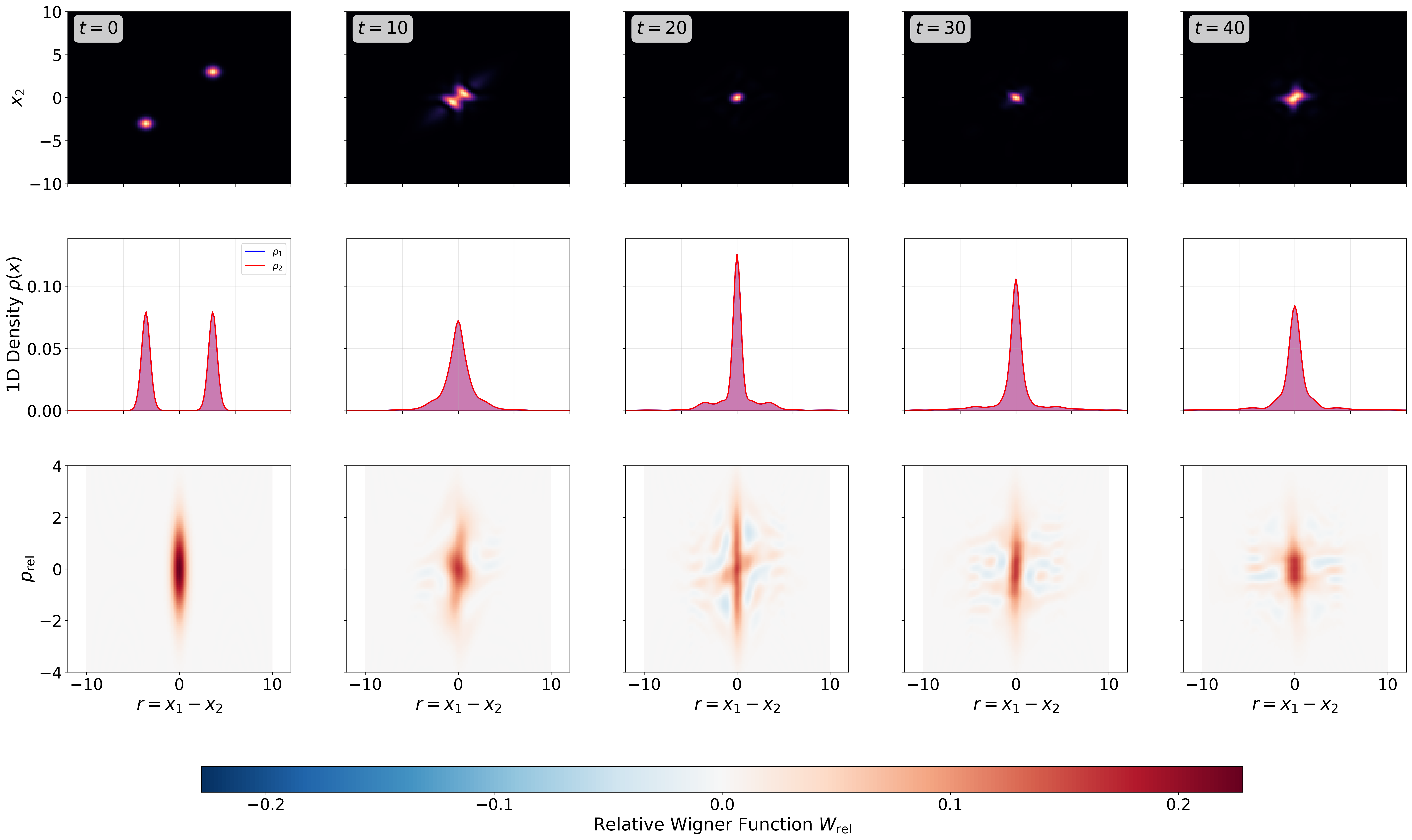}
\caption{
Evolution of the correlated superposition state $\Psi_{\rm III}$ (Eq.~\ref{eq:psi_III})
using stationary SN profiles.
The entanglement entropy oscillates around $S_{\rm vN}\approx 1.28$,
reflecting deformation of the two same-side components under the common self-field and pair
attraction. The panel shows (top row) the joint probability density $|\Psi(x_1,x_2)|^2$;
(middle row) single-particle marginal densities $\rho_1(x)$ and $\rho_2(x)$; (bottom row)
the relative-coordinate Wigner function $W_{\rm rel}$, with snapshots at $t=0, 10, 20, 30,
40$.
Parameters: $R_0=6$, $\kappa=\gamma=1$; the imaginary-time seed width is $\sigma_0=1$.
}
\label{fig:collision_correlated}
\end{figure}

The gravitational self-interaction acts symmetrically on the two components
(Fig.~\ref{fig:collision_correlated}), and the entanglement entropy
oscillates around $S_{\rm vN}\approx 1.28$ without systematic growth or
decay. The marginal center-of-mass separation $|\langle x_1 \rangle - \langle x_2
\rangle|$ remains near zero throughout the evolution by symmetry. However, examining the
local density reveals non-stationary dynamics driven by the non-linearity of the SN
equation. Because the self-potential $\Phi(x)$ is sourced by the marginal density, which
is split equally between $L$ and $R$ (for negligible overlap $s$), the local part of the self-gravitational well at each branch is approximately
half as deep as for an isolated stationary profile. The original profile is therefore
no longer a stationary solution of this weakened potential and undergoes immediate kinetic
dispersion. Simultaneously, the density of the remote spatial component generates a
secondary gravitational attraction toward the center, accelerating the components into a
collision at the origin.
The entanglement rises from its initial value (close to $\ln 2$ for the numerically small
overlap used here) to oscillate around $S_{\rm vN} \approx 1.28$. Because this value exceeds the rigid-profile maximum of $\ln 2 \approx 0.69$, the full pair-interaction dynamics clearly excites higher-order spatial modes beyond the initial left/right subspace.

In phase space, both particles are always co-localized ($r \approx 0$), so the relative
Wigner function starts as a single localized lobe at the origin with no macroscopic
interference fringes. Because the relative coordinate is already centered, the mutual
gravitational force does not drive macroscopic transport in $r$. The state undergoes local
squeezing as the marginals collapse, which explains why the entanglement oscillates at a
moderate level without systematic growth.

\subsubsection{Anticorrelated superposition}
\label{sec:anti_collision}

The anticorrelated configuration $\Psi_{\rm IV}$ (Eq.~\ref{eq:psi_IV}) places the
particles on opposite sides. This state is also entangled at $t=0$, but unlike the
correlated case, its two components ($L,R$ and $R,L$) are spatially disjoint.

\begin{figure}[htbp]
\centering
\includegraphics[width=\linewidth]{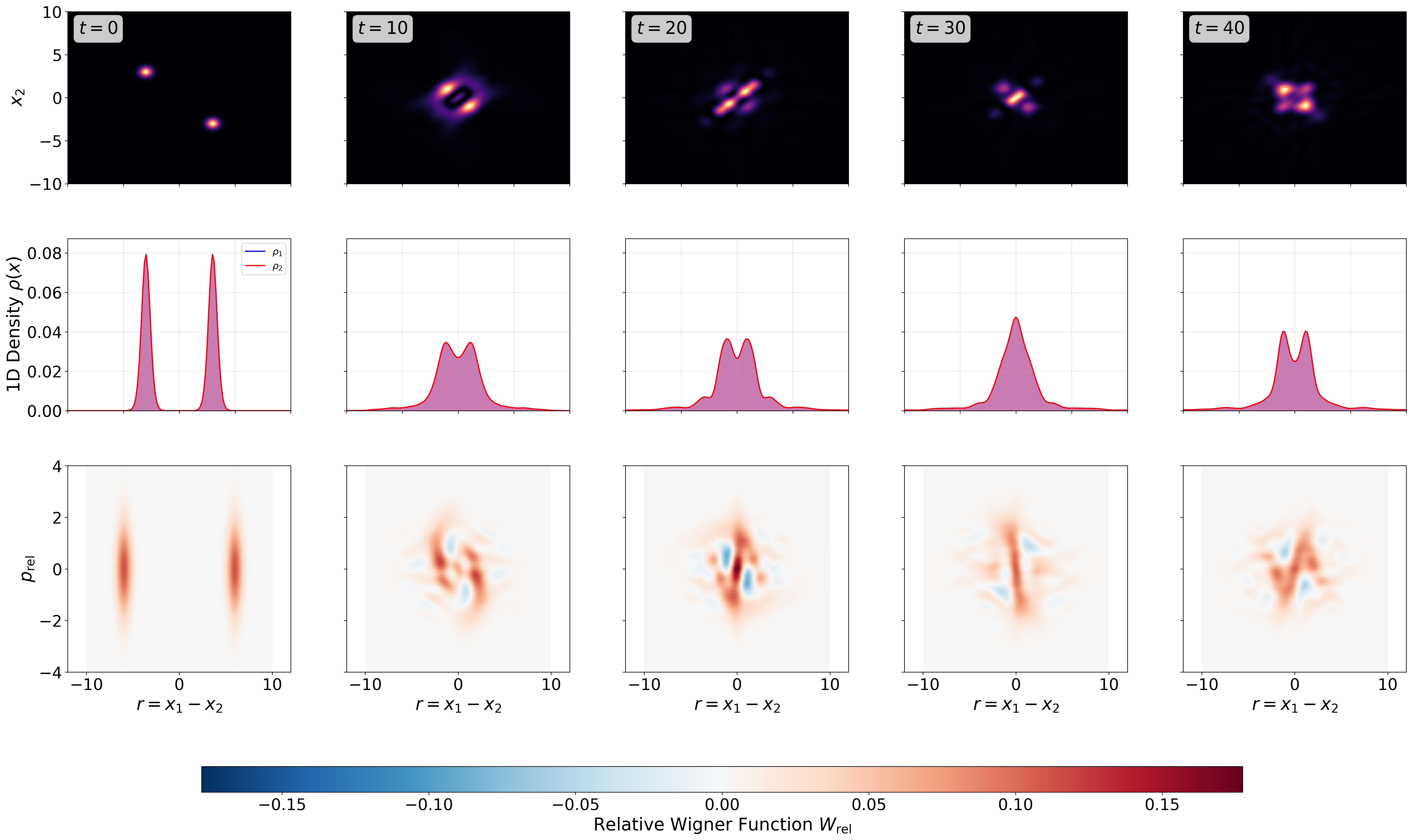}
\caption{
Evolution of the anticorrelated superposition state $\Psi_{\rm IV}$ (Eq.~\ref{eq:psi_IV})
using stationary SN profiles.
The mutual gravitational attraction drives the disjoint spatial components
together, increasing the entanglement entropy to $S_{\rm vN}\approx 1.67$. The panel shows
(top row) the joint probability density $|\Psi(x_1,x_2)|^2$; (middle row) single-particle
marginal densities $\rho_1(x)$ and $\rho_2(x)$; (bottom row) the relative-coordinate
Wigner function $W_{\rm rel}$, with snapshots at $t=0, 10, 20, 30, 40$.
Parameters: $R_0=6$, $\kappa=\gamma=1$; the imaginary-time seed width is $\sigma_0=1$.
}
\label{fig:collision_anticorrelated}
\end{figure}

In the anticorrelated superposition (Fig.~\ref{fig:collision_anticorrelated}), the mutual
gravitational attraction actively pulls the particles in each component
toward each other, driving the $r=\pm R_0$ relative-coordinate components toward overlap
near $r=0$.
The entanglement entropy reaches $S_{\rm vN}\approx 1.67$, comparable
to the delocalized product, and the rms relative separation $D_{\rm rel}(t)$ undergoes
large-amplitude oscillations as the components interfere constructively
and destructively during their gravitational evolution.

The relative coordinate is in a macroscopic superposition of $r = \pm R_0$. At $t=0$, the
relative Wigner function has two localized lobes at $r = \pm R_0$ separated by
high-frequency interference fringes. The mutual gravitational force accelerates these
components toward each other, shearing them in phase space. Their collision at the origin
produces Wigner negativity.

\subsubsection{Entanglement dynamics}
\label{sec:comparison}

\begin{figure*}[htbp]
\centering
\includegraphics[width=0.85\linewidth]{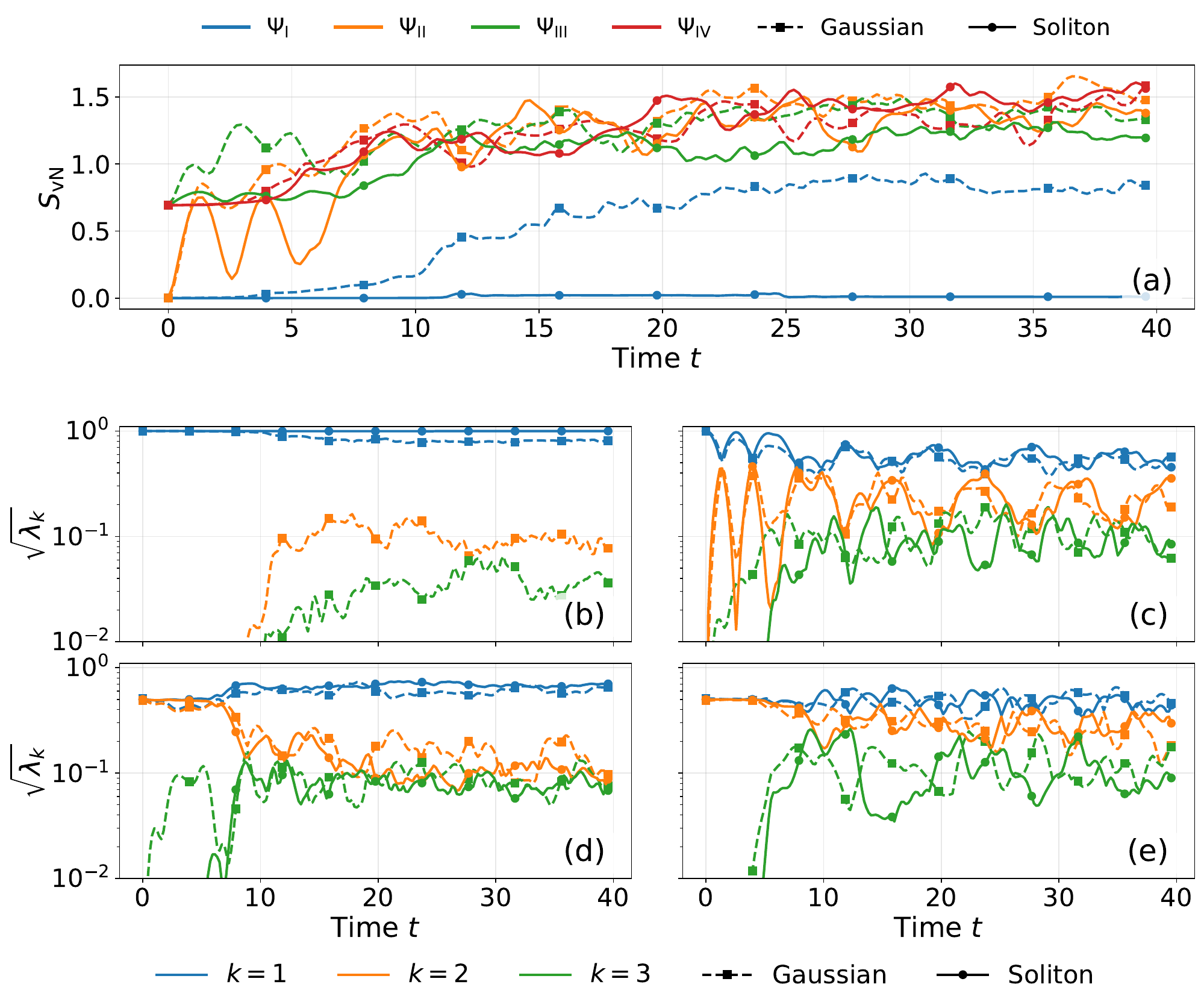}
\caption{
Entanglement and Schmidt spectrum dynamics comparing Gaussian wavepackets (solid lines)
and stationary SN profiles (dashed lines with circular markers) across the four initial
condition configurations. (a) Evolution of the von Neumann entanglement entropy $S_{\rm
vN}$. (b-e) Evolution of the Schmidt spectrum (eigenvalues $\lambda_k$) for the lowest
three modes: (b) Localized product $\Psi_{\rm I}$ (Eq.~\eqref{eq:psi_I}), (c) Delocalized product $\Psi_{\rm
II}$ (Eq.~\eqref{eq:psi_II}), (d) Correlated superposition $\Psi_{\rm III}$ (Eq.~\eqref{eq:psi_III}), and (e) Anticorrelated superposition
$\Psi_{\rm IV}$ (Eq.~\eqref{eq:psi_IV}). In all cases, stationary profiles suppress the transfer of weight into
higher Schmidt modes compared with Gaussian packets, where kinetic spreading and
collision-induced distortion accelerate Schmidt-spectrum broadening.
}
\label{fig:schmidt_grid}
\end{figure*}

Figure~\ref{fig:schmidt_grid} compares $S_{\rm vN}(t)$ and the leading Schmidt
eigenvalues for Gaussian and stationary-profile initial data. Because the self-field is isospectral, all spectrum changes are due to $V_{\rm pair}$.

Among stationary-profile simulations, the localized product produces the smallest
dynamically generated entropy, reaching $S_{\rm vN}\approx 0.19$ over the plotted window.
The correlated cat state starts close to $\ln 2$ and oscillates around $S_{\rm vN}\approx
1.2$, while the anticorrelated and delocalized product states reach the largest absolute
entropies ($S_{\rm vN}\approx 1.58$ and $1.67$), driven by the different pair-potential phases and
accelerations sampled by co-localized and separated components. For the parameters studied
here, the entropy hierarchy is controlled mainly by spatial geometry: states with multiple
relative-coordinate components populate higher Schmidt modes more efficiently than localized
product collisions.

Figure~\ref{fig:schmidt_grid}(b-e) shows the corresponding redistribution of Schmidt
weight. The decay of $\lambda_1$ and the growth of $\lambda_2,\lambda_3$ indicate
occupation of spatial modes beyond the initial left/right subspace. Gaussian wavepackets
show stronger redistribution because kinetic spreading increases the spatial support over
which the pair potential varies. For stationary profiles, self-localization keeps the
packets narrower and suppresses this transfer to higher Schmidt modes.

\subsection{Mass asymmetry}
\label{sec:mass_asymmetry}

We next consider unequal masses, keeping $\mu_2=1$ fixed and increasing $\mu_1$. We
consider a localized-product collision with a mass ratio of $4{:}1$ ($\mu_1 = 4.0$, $\mu_2
= 1.0$), using the universal coupling $\kappa=\gamma=1.0$.

\begin{figure*}[htbp]
\centering
\includegraphics[width=\linewidth]{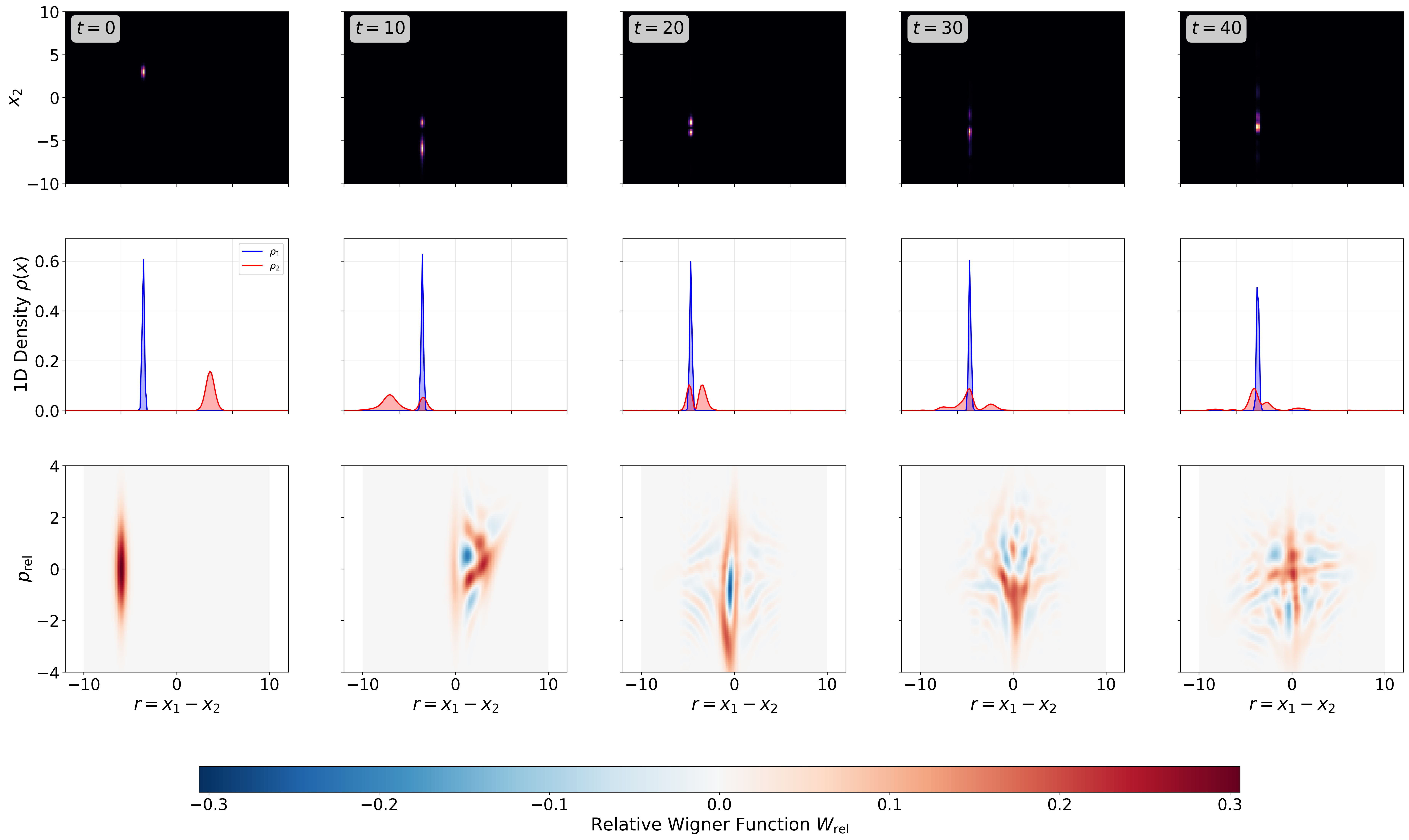}
\caption{
Mass-asymmetric localized-product collision with $\mu_1=4.0$ and $\mu_2=1.0$. The heavy
particle (blue) remains nearly stationary while the light particle (red) is accelerated
and deformed during the collision. The panel shows (top row) the joint probability density
$|\Psi(x_1,x_2)|^2$; (middle row) single-particle marginal densities $\rho_1(x)$ and
$\rho_2(x)$; (bottom row) the relative-coordinate Wigner function $W_{\rm rel}$, with
snapshots at $t=0, 10, 20, 30, 40$. Parameters: $R_0=6$, $\kappa=\gamma=1$; the
imaginary-time seed width is $\sigma_0=1$.
}
\label{fig:mass_asymmetry_snapshots}
\end{figure*}

Figure~\ref{fig:mass_asymmetry_snapshots} shows the joint density, marginals, and relative
Wigner function for this mass-asymmetric collision. Owing to the steep $\mu^2$ scaling of
the self-gravitational well depth, the heavy stationary profile becomes narrow on the
scale of the light profile and acts approximately as a localized scattering potential. By
contrast, the lighter particle has a broader stationary profile with weaker self-binding.
As the mutual attraction drives the packets together, the broad light profile scatters
from this narrow heavy-mass potential.

Because the light particle's self-binding ($\kappa\mu_2^2=1.0$) is weaker than that of the
heavy particle, the light stationary profile is deformed during the collision. The heavy
profile acts as a sharp scattering center, exciting oscillatory distortions of the lighter
marginal density. This deformation is visible in the relative Wigner function (bottom row
of Fig.~\ref{fig:mass_asymmetry_snapshots}), which develops interference fringes near the
collision. The corresponding entropy increase reaches $S_{\rm vN}\approx
1.94$, indicating that large-scale fragmentation at mass
ratio $4{:}1$ produces substantial entanglement.

To quantify the broadening of the lighter marginal density, we use the Participation Ratio
\begin{equation}
\mathrm{PR} = \left[ \int \rho^2(x)\,dx \right]^{-1},
\label{eq:pr}
\end{equation}
which scales
linearly with the spatial extent of the wavepacket. Unlike the variance, the PR weights
the high-density support of the marginal and is less sensitive to small tails at large
$|x|$.

\begin{figure*}[htbp]
\centering
\includegraphics[width=\linewidth]{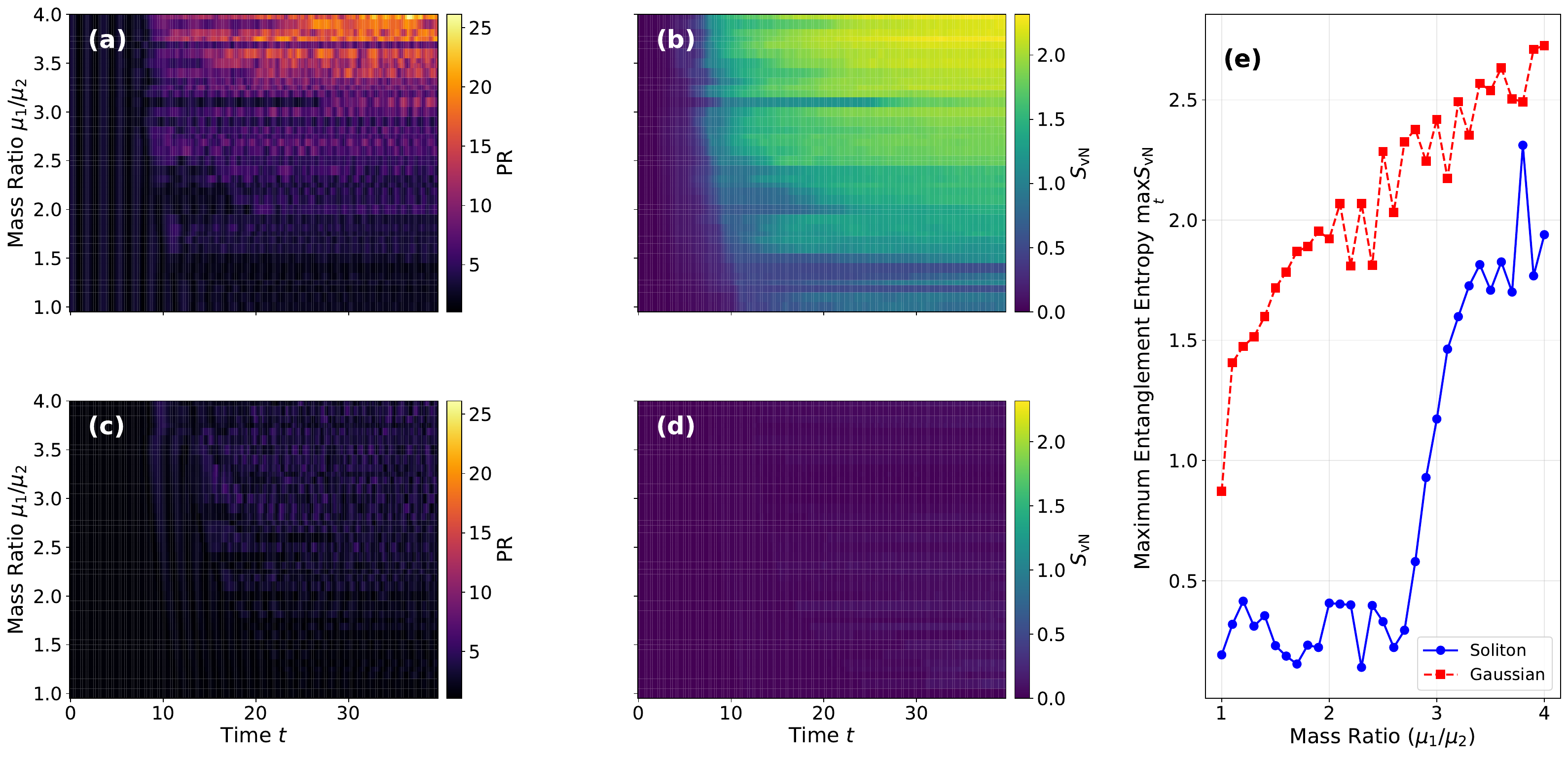}
\caption{
Gravitational shattering and entanglement generation in mass-asymmetric collisions.
(a,c) Heatmaps of the Participation Ratio [PR, Eq.~\eqref{eq:pr}] of the lighter
particle's marginal density $\rho_2(x,t)$ as a function of time and mass ratio $\mu_1/\mu_2$ for
Gaussian wavepackets (a) and stationary SN profiles (c). Spatial deformation appears as an
increase in PR. The dispersing Gaussian packet (a) undergoes significant shattering, whereas
the stationary profile (c) suppresses deformation.
(b,d) Corresponding heatmaps of the von Neumann entanglement entropy $S_{\rm vN}(t)$ for
Gaussian (b) and stationary (d) initial states. Entanglement generation strongly correlates
with the spatial disruption of the lighter mass.
(e) Peak entanglement $\max_t S_{\rm vN}$ (over $t \in [0, 40]$) versus mass ratio $\mu_1/\mu_2$.
Stationary-profile entanglement remains minimal and nearly constant across mass ratios, while
Gaussian entanglement grows rapidly as the scattering potential acts on the dispersing,
fragmented wavepacket.
}
\label{fig:mass_asymmetry_combined}
\end{figure*}

Figure~\ref{fig:mass_asymmetry_combined} (Left) maps the PR of the lighter particle's
marginal density as a function of time and mass ratio. Increases in PR indicate spatial
broadening and deformation of the light particle. For stationary SN profiles this
broadening remains comparatively limited over the scanned range, whereas Gaussian packets
display a much more persistent PR increase because kinetic dispersion exposes them to
larger regions of the scattering potential.

For stationary SN profiles, self-binding limits the PR growth and the associated
Schmidt-spectrum broadening; for Gaussian packets, prior kinetic spreading increases the
spatial region over which the tidal gradient acts.

Figure~\ref{fig:mass_asymmetry_combined} (Right) shows the peak entanglement $\max_t
S_{\rm vN}$ as a function of the mass ratio $\mu_1/\mu_2$, with $\mu_2=1.0$ held fixed,
for both Gaussian wavepackets and stationary SN profiles.

For stationary SN profiles, the peak entanglement remains nearly constant across the
scanned mass ratios: self-binding suppresses spatial fragmentation during the collision.
The freely dispersing Gaussian wavepackets, by contrast, undergo increasingly pronounced
spatial disruption by the heavy mass, and the entanglement entropy grows steadily with
$\mu_1/\mu_2$.

\section{Conclusions and Outlook}
\label{sec:conclusions}

The question of entanglement generation through gravity is of primary interest in  foundational studies \cite{QG_RMP}. It has been claimed that classical gravity cannot generate entanglement between spacelike separated particles and, consequently, that the observation of gravitationally-induced entanglement would provide a decisive argument for the quantumness of the gravitational field (see \cite{Belenchia2018,Aurell21,Selby22,Marletto25}). However, this argument hinges upon the assumption that the gravitational interaction is \textit{local}. This can be questioned, given the fact there exist no local expression for the energy of the gravitational field \cite{Szabados09}. While the single-particle SN equation is not local in the spacetime sense \cite{Selby22,Osekat2025}, its canonical 2-particle version \cite{Diosi1984} \emph{is} local in the sense of bipartite dynamics \cite{Paterek2024}. In Appendix \ref{sec:appendix_hartree} we show that this is generally the case in models admitting the Hartree projection onto the product-state manifold.

We analyzed a two-body Schr\"odinger--Newton-type model that explicitly separates gravitational entanglement generation from mean-field self-localization. The numerical analysis demonstrates that entanglement growth depends sensitively on spatial dispersion. Highly localized, self-bound profiles undergo quasi-elastic scattering with minimal Schmidt-spectrum excitation. Kinetically dispersing wavepackets amplify the entangling power of the pair potential. In asymmetric collisions, the gravitational field of the heavier mass spatially disrupts the lighter wavepacket, producing pronounced Wigner negativity and rapid entanglement growth. Consequently, any semiclassical formulation combining Di\'osi--Penrose self-localization with Newtonian mutual attraction must explicitly define whether the interaction preserves separability as a mean field or operates as a nonseparable pair interaction.

The most direct extensions are 2D and 3D scattering geometries with finite impact parameter, $N$-body collisions, and the addition of an external disorder potential to study the interplay with Anderson-type localization. Furthermore, developing realistic physical models will require moving beyond these simplified low-dimensional head-on collisions to full 3D simulations. Such extensions are crucial for determining whether this direct gravitational entanglement generation could ever be isolated and measured in next-generation optomechanical levitation experiments.

\section*{Acknowledgments}
MP acknowledges funds from MICIU/AEI/10.13039/501100011033/ FEDER, UE. M.L. acknowledges support from: European Research Council AdG NOQIA; MCIN/AEI (PGC2018-0910.13039/501100011033, CEX2019-000910-S/10.13039/501100011033, Plan National FIDEUA PID2019-106901GB-I00, Plan National STAMEENA PID2022-139099NB, I00, project funded by MCIN/AEI/10.13039/501100011033 and by the “European Union NextGenerationEU/PRTR" (PRTR-C17.I1), FPI); QUANTERA DYNAMITE PCI2022-132919, QuantERA II Programme co-funded by European Union’s Horizon 2020 program under Grant Agreement No 101017733; Ministry for Digital Transformation and of Civil Service of the Spanish Government through the QUANTUM ENIA project call - Quantum Spain project, and by the European Union through the Recovery, Transformation and Resilience Plan - NextGenerationEU within the framework of the Digital Spain 2026 Agenda;
Fundació Cellex; Fundació Mir-Puig;
Generalitat de Catalunya (European Social Fund FEDER and CERCA program; Barcelona Supercomputing Center MareNostrum (FI-2023-3-0024); Funded by the European Union. Views and opinions expressed are however those of the author(s) only and do not necessarily reflect those of the European Union, European Commission, European Climate, Infrastructure and Environment Executive Agency (CINEA), or any other granting authority. Neither the European Union nor any granting authority can be held responsible for them (HORIZON-CL4-2022-QUANTUM-02-SGA PASQuanS2.1, 101113690, EU Horizon 2020 FET-OPEN OPTOlogic, Grant No 899794, QU-ATTO, 101168628), EU Horizon Europe Program (This project has received funding from the European Union’s Horizon Europe research and innovation program under grant agreement No 101080086 NeQSTGrant Agreement 101080086 — NeQST); ICFO Internal “QuantumGaudi” project;
J.O. and M.E. acknowledge support from the National Science Centre in Poland under the research grant Sonata BIS \textit{Beyond Quantum Gravity} (2023/50/E/ST2/00472).
\appendix

\section{Hartree reduction and recovery of mean-field models}
\label{sec:appendix_hartree}

The separability-preserving Schr\"odinger--Newton formulation studied in
Ref.~\cite{Paterek2024} is characterized by gravitational interactions that preserve the
product form of initially product many-body states. Within the present variational
hierarchy, the same separability-preserving structure is obtained as the Hartree
projection of the pair-interaction dynamics onto the product-state manifold via the
Dirac--Frenkel variational principle~\cite{Dirac1930,Frenkel1934}.

We impose the Hartree product ansatz on the pair-interaction model:
\begin{equation}
\Psi(x_1,x_2,t)=\psi_1(x_1,t)\psi_2(x_2,t),
\label{eq:appendix_hartree_ansatz}
\end{equation}
with marginals $\rho_i(x_i,t)=|\psi_i(x_i,t)|^2$ normalized to unity. To derive the
mean-field limit, we constrain the system to the Hartree manifold---the separable
submanifold of $\mathcal{H}_1 \otimes \mathcal{H}_2$ spanned by exact product states.
Rather than linearizing the SN self-field, we restrict the full energy functional
associated with Eq.~\eqref{eq:pair_SN} to this zero-entanglement geometry. For a
factorized state \(\Psi=\psi_1\psi_2\), the functional decomposes as
\begin{equation}
E[\psi_1,\psi_2]=E_1[\psi_1]+E_2[\psi_2]+E_{\rm pair}[\psi_1,\psi_2],
\end{equation}

Because the state is an exact product, the kinetic and self-energy terms decouple
trivially, yielding local energy terms for each particle:
\begin{equation}
E_1[\psi_1] = \int dx_1\, \psi_1^*(x_1,t) \left( -\frac{1}{2\mu_1}\partial_{x_1}^2 \right) \psi_1(x_1,t) 
- \frac{\kappa}{2}\mu_1^2 \iint dx_1\,dx_1'\, |\psi_1(x_1,t)|^2 U_\epsilon(x_1-x_1') |\psi_1(x_1',t)|^2,
\label{eq:E1_functional}
\end{equation}
and similarly for $E_2[\psi_2]$. The factor of $1/2$ prevents double-counting of the self-energy. 
The expectation value of the nonseparable pair interaction $V_{\rm pair}$ evaluated on
this product manifold reduces to the density-density interaction energy between the two
marginal densities:
\begin{equation}
E_{\rm pair}[\psi_1, \psi_2]
=
-\gamma\mu_1\mu_2
\int dx_1\,dx_2\,
|\psi_1(x_1,t)|^2|\psi_2(x_2,t)|^2
U_\epsilon(x_1-x_2).
\label{eq:Epair_functional}
\end{equation}

To obtain the equations of motion for the individual factors $\psi_1$ and $\psi_2$, we
apply the Dirac--Frenkel--McLachlan time-dependent variational
principle~\cite{Dirac1930,Frenkel1934,McLachlan1964} to the action
\begin{equation}
\mathcal{S}[\Psi] = \int dt\, \left[ i\int dx_1\,dx_2\,\Psi^*\partial_t\Psi - E[\Psi] \right],
\end{equation}
where $E[\Psi]$ is the conserved energy functional. This formulation is essential for the
nonlinear SN terms: using the expectation value of the instantaneous effective Hamiltonian
would double-count the self-gravitational energy. For the restricted product-state
manifold, we demand stationarity under independent variations $\delta\psi_1^*$ and
$\delta\psi_2^*$.

For a normalized product state $\Psi = \psi_1\psi_2$, the expectation value of the
time-derivative operator expands as:
\begin{equation}
\int dx_1\,dx_2\, \Psi^* i\partial_t \Psi = i \int dx_1\, \psi_1^* \partial_t \psi_1 + i \int dx_2\, \psi_2^* \partial_t \psi_2.
\end{equation}
The variations are taken tangent to the normalized Hartree manifold; equivalently,
one may introduce time-dependent Lagrange multipliers enforcing
$\|\psi_1\|=\|\psi_2\|=1$, which only modify the gauge terms below.
Setting the variational derivative of the action with respect to $\psi_1^*(x_1)$ to zero yields:
\begin{equation}
i\,\partial_t\psi_1(x_1,t) + i \left(\int dx_2\,\psi_2^*\partial_t\psi_2\right) \psi_1(x_1,t) = \frac{\delta}{\delta\psi_1^*(x_1,t)} E[\psi_1, \psi_2].
\label{eq:appendix_df_var}
\end{equation}
We evaluate the functional derivatives $\delta E / \delta\psi_1^*$ term by term. For the
single-particle functional $E_1$, the variation with respect to $\psi_1^*$ yields the
standard nonlinear Schr\"odinger--Newton term, where the factor of $1/2$ is cancelled by
the product rule applied to the quartic interaction $|\psi_1|^2|\psi_1'|^2$:
\begin{align}
\frac{\delta E_1}{\delta\psi_1^*(x_1,t)} &= \left[ -\frac{1}{2\mu_1}\partial_{x_1}^2 - \kappa\mu_1^2 \int U_\epsilon(x_1-x_1')|\psi_1(x_1',t)|^2\,dx_1' \right] \psi_1(x_1,t) \nonumber \\
&= \left[ -\frac{1}{2\mu_1}\partial_{x_1}^2 - \kappa\mu_1\Phi_1(x_1,t) \right] \psi_1(x_1,t).
\end{align}

Next, taking the variational derivative of the pair energy $E_{\rm pair}$ with respect to
$\psi_1^*$ yields an effective one-body potential generated by the partner's density:
\begin{equation}
\frac{\delta E_{\rm pair}}{\delta\psi_1^*(x_1,t)} = \left[ -\gamma\mu_1\mu_2 \int U_\epsilon(x_1-x_2)|\psi_2(x_2,t)|^2\,dx_2 \right] \psi_1(x_1,t).
\end{equation}
We identify the integral bracket as the effective mean-field cross-coupling:
\begin{equation}
\widetilde{\Phi}_{12}^{(2\to1)}(x_1,t)
=
\int U_\epsilon(x_1-x_2)\rho_2(x_2,t)\,dx_2.
\end{equation}
By symmetry, the corresponding variations with respect to $\psi_2^*$ yield the analogous
terms for the second particle, including the reciprocal mean-field potential:
\begin{equation}
\widetilde{\Phi}_{12}^{(1\to2)}(x_2,t) = \int U_\epsilon(x_1-x_2)\rho_1(x_1,t)\,dx_1.
\end{equation}

Combining these variations, the resulting Hartree-projected equations of motion are:
\begin{align}
i\,\partial_t\psi_1(x_1,t)
&=
\left[
-\frac{1}{2\mu_1}\partial_{x_1}^2
-\kappa\mu_1\Phi_1(x_1,t)
-\gamma\mu_1\mu_2\widetilde{\Phi}_{12}^{(2\to1)}(x_1,t)
\right]\psi_1(x_1,t) + C_1(t)\psi_1(x_1,t),
\label{eq:appendix_hartree_1} \\
i\,\partial_t\psi_2(x_2,t)
&=
\left[
-\frac{1}{2\mu_2}\partial_{x_2}^2
-\kappa\mu_2\Phi_2(x_2,t)
-\gamma\mu_1\mu_2\widetilde{\Phi}_{12}^{(1\to2)}(x_2,t)
\right]\psi_2(x_2,t) + C_2(t)\psi_2(x_2,t),
\label{eq:appendix_hartree_2}
\end{align}
where the purely time-dependent phase constants are $C_1(t) = -i\left(\int dx_2\,\psi_2^* \partial_t \psi_2 \right)$ and $C_2(t) = -i\left(\int dx_1\,\psi_1^* \partial_t \psi_1 \right)$.
These are pure gauge terms that can be set to zero by imposing the parallel-transport
conditions $\int dx_i\,\psi_i^*\partial_t\psi_i=0$.

Equations~\eqref{eq:appendix_hartree_1}--\eqref{eq:appendix_hartree_2} have the same
separability-preserving Hartree structure as the Schr\"odinger--Newton equations of
Ref.~\cite{Paterek2024}:
\begin{align}
i\,\partial_t\psi_1 &= \left[T_1 + V_{11}[|\psi_1|^2] + V_{12}[|\psi_2|^2] \right]\psi_1, \\
i\,\partial_t\psi_2 &= \left[T_2 + V_{22}[|\psi_2|^2] + V_{21}[|\psi_1|^2] \right]\psi_2,
\end{align}
where $V_{ii}$ are self-potentials and $V_{ij}$ are mean-field cross-potentials.
Thus, separability-preserving SN models arise naturally as the Hartree projection of the
pair-interaction dynamics. This product reduction precludes entanglement generation by
construction. For a universal coupling $\kappa=\gamma\equiv\lambda$, these projected
equations match the product-state restriction of a semiclassical mean-field model where gravity is
sourced by the expectation value $\langle\hat{\rho}\rangle$, up to the chosen dimensional
reduction.

\bibliographystyle{naturemag}
\bibliography{references}

\end{document}